\documentclass{article}
\usepackage{graphicx} 

\title{CONTEMPORARY COSMOLOGY FROM LAKATOS' VIEWPOINT}
\author{J.E. Horvath (IAG-USP, S\~ao Paulo, Brazil)}
\date{foton@iag.usp.br}

\begin{document}

\maketitle
ABSTRACT: An analysis of contemporary Cosmology is presented, with the aim of identifying the elements present in it according to the scientific program structure created by I. Lakatos. We look at some modern controversies from this point of view and clarify the meaning of issues related to them within this context.


\section{Introduction}

The contemporary picture of Cosmology has been slowly developed over the 20th century and the first decades of the 21st, on a quite successful path which featured theoretical insight and observational work prompted by increasingly sophisticated instruments and techniques. This modern phase started with the construction of relativistic Cosmology, based on General Relativity, and received input from other areas such as Nuclear and Particle Physics, once the idea of a hot and dense early Universe (the Big Bang) was put forward and seriously developed. 
Within this perspective, we arrived to the present paradigm, the so-called $\Lambda$CDM (Lambda Cold Dark Matter) Cosmology. For an outsider, it is hard to understand why cosmologists should be satisfied: the cold dark matter is an undetected component thought to be fundamental for structure formation and largely dominant in many scales. On the other hand, a (constant) energy density $\Lambda$ is the latest ingredient of this model, an unclustered form of invisible energy making the Universe expansion to accelerate, and adding up to ~ 70$\%$ of the full matter/energy content of the Cosmos. These two entities have been indirectly inferred, and there are big observational/experimental efforts to detect some direct evidence, and also a huge theoretical activity to go beyond mere "coincidence Cosmology" and justify a complete picture of the Universe. 

In addition to this disciplinary viewpoint, the philosophical constructions of the 20th century, championed by Popper, Kuhn, Lakatos, Feyerabend and others can be invoked to visualize the whole picture and its possible future path. We have argued before (Horvath 2009) that the acceleration of the Universe triggered a kind of "extraordinary science" period, such as suggested by T. Kuhn (1962). The "anomaly" evidenced by the acceleration, which is now 20+ years old, may or may not be solved by establishing a Dark Energy or otherwise. Meanwhile, the other "anomaly" (the lack of material to match the observation, leading to the Dark Matter hypothesis) will be a century old soon (Zwicky 1933), and nothing convincing has been detected to play this role (Shutt 2013), and also theoretical suggestions are many and never reached any firm consensus (Oks 2021). Moreover, other controversies ("anomalies" in Kuhn's terminology) have arisen and may be important for the future of the field. Therefore, it is interesting to go back to the 20th century ideas and revisit the question of how is Cosmology constructed as a research programme, identifying the main ingredients and putting issues in perspective. This is precisely the task we will attempt here, using the framework developed by Lakatos in the last century, with the aim of visualizing a more clear picture of the whole discipline and perhaps giving some insight on its future development.

\section{The hard core and protective belt of modern Cosmology}

In Lakatos' view, the present $\Lambda$CDM Cosmology can be considered a paradigm (Kuhn 1962), basically a research programme that became hegemonic (Lakatos 1978). Most cosmologists believe, even with some evidence against it, that $\Lambda$CDM Cosmology is essentially correct, and will not open any discussion to confront it with any other model, for example, the now abandoned Steady State (Kragh 1999). 
$\Lambda$CDM Cosmology is the last version of 20th century Cosmology, which evolved from a philosophical/speculative state towards a real empirical science over the years, mainly after the Hubble announcement of the expansion of the Universe, in which Slipher, Humason, Lema{\^i}tre and others contributed. This discovery shifted the interest from a static, infinitely old Universe to a dynamical, finite age one.
By its very construction and history, the $\Lambda$CDM Cosmology shows different elements within, which evolved individually and as a intertwined set. An appraisal of these constitutive elements follow below. 

\subsection{The hard core}

Over the years, and after the consolidation of the Einstenian programme for the description of gravity, a few fundamental elements constituting the hard core of the Cosmology research programme can be identified. These elements are never disputed, and if some problem arises with them, the protective belt can be modified to save their validity (see below). We suggest that the hard core of modern Cosmology is constituted by

$\ast$ The Cosmological Principle

A fundamental (meta)postulate in Cosmology has its origins in the revolutionary work of  N. Copernicus around 1500 A.D., which rediscovered and extended the observations of Aristarchus of Samos (III century B.C.). There has been a lot of discussion on the true extent of Copernicus' work, mainly about the consequences of formulating an heliocentric model that removed the Earth from the center of the Cosmos. 

The basic Copernican Principle is usually stated as there are no "special" observers. This sharply contrasts with, for example, Aristotelian Cosmology, in which the Earth occupies a privileged position in the whole world system. Sometimes it has been insisted that the inference that we are "cosmically ordinary" should be called the principle of Copernican mediocrity (with some subtle differences not discussed here). And as a deeper scrutinized issue, the elaborated Genuine Copernican Cosmological Principle which says: The Universe as observed from any planet looks much the same. The latter form is specially suitable to link Copernicus with the modern form of the Cosmological Principle, which has been argued to be substantially stronger than the former (Beisbart \& Jung 2006): Each observer in the Universe, in any position, would observe the same patterns independently of the direction. 
A Universe which has no privileged place (center or differentiated places for the observers), and which shows the same large-scale pattern in any direction is called homogeneous and isotropic in Cosmology jargoon. We see that statements based on the observer's situation and thus translated into statements about the properties of the Universe itself. Mathematical models should possess these properties to comply with the postulated observer status. This is indeed the case of  $Lambda$CDM Cosmology, and challenges that have arisen by observations which seem at odds with the Cosmological Principle (see below).

$\ast$ Conservation of energy

The conservation of energy has been for a very long time a metapostulate in Physics, and later related to time-symmetric formal mathematical descriptions via Noether's currents (Baez 2020). Ideas suggesting a violation of energy conservation were suggested from time to time, especially when new branches or phenomena emerged (Landau 1932), but refuted empirically (Bothe \& Geiger 1925), and indeed found  not viable by most of the practitioners. This energy conservation certainly applies to any local process.

It is then easy to envisage why (besides several lines of evidence in favor of an expanding Cosmos) the Steady State theory was progressively abandoned: to maintain a steady Universe, a creation field (C-field) had to be introduced, and related to the Faraday-Lenz phenomenon because the creation of matter/energy opposes the Universe dynamical state. This C-field can be interpreted as creating matter/energy from the vacuum, and therefore violating the simplest form of energy conservation. However, there is an argument stating that the Universe is not time-invariant (it changes over time irreversibly), and therefore it does not conserve energy as a whole, since the latter is the charge associated with time symmetry, although many do not consider that this is enough to consider a C-field. General Relativity and most viable alternatives do not rely on any C-field, a feature which is considered a desirable feature by cosmologists. The community welcomes the solution of cosmological problems while conserving energy, elevated to the range of hard core postulate.

It is amusing to consider that the "modern" form of this conservation law is actually the last version of Lucretius statement "nothing can be made from nothing", initially expressed by Parmenides (Burnet 2014) many centuries before him. While "energy" is the contemporary form of the conserved quantity (we now think that mass is just a concentrated form of it), the idea is the basically same. It can be said that this element of the Cosmology hard core programme is as old as the Western thought itself.

$\ast$ Metric character of gravity

The last suggested element of the hard core is more technical, and related to the identification of gravitation as a deformation of spacetime, the vision championed by Einstein. The idea here is that the fundamental equations of motion describing gravitation relate the sources (matter and energy as a whole) to the deformation of spacetime, described by a function that essentially measures distances between spacetime events. This mathematical object is called a metric, and implies that the mathematical space in which it lives is endowed with a consistent definite expression to measure distance between events happening in spacetime (i.e. it is a metric space). General Relativity is the prime example of such a class, but there is a large family of metric theories which share this underlying structure, and many alternatives to General Relativity belong to this class. Different hypothesis separate the latter alternative theories, for example, Unimodular Gravity is very similar to General Relativity, but an extra condition on the former can be interpreted as giving rise to a non-zero cosmological constant. Many other alternatives have been suggested (Bellucci, Faraoni \& Longo, 2022), including some non-metric ones, but the success of General Relativity in many tests has created a kind of consensus that, if some alternative is presented, it can hardly escape a metric character.

An important battery of tests that have not been fully explored yet is provided by the decade-old precise measurements (Kramer et al. 2021) of the double pulsar system, in which two pulsating neutron stars have provided a way to test the Post-Newtonian and some Post-Post-Newtonian corrections deviations from General Relativity. These deviations have proved to be $\leq \, 0.01 \%$ at most, in the strong gravity regime, implying that General Relativity leaves little room for alternative theories. However, this fact will not be meaningful for Cosmology until alternative cosmologies (including those based in non-metric schemes) can be subject to analogue tests and confirm their status. This is a research programme that has not really started and can be reveling for the progress in Cosmology.
It is important to remark that the hard core content underlies the formal mathematical structure of $\Lambda$CDM Cosmology, and suffices for its full formulation. For example, acceptance of the Cosmological Principle and energy conservation will produce the Friedmann-Lema{\^i}tre-Robertson-Walker metric as the only possibility within General Relativity. Contemporary cosmologists who believe that gravitation has a dynamical metric character would seek for modified FLRW equations (see below) and postulate DM/DE added to the matter/energy content or a description of gravitation that preserves the general feature of a metric. All these components/modifications will be attached to the protective belt, specifically to the negative heuristics (see below) in Lakatos' terminology.

\begin{figure}[htbp]
\centering
\includegraphics[width=10cm]{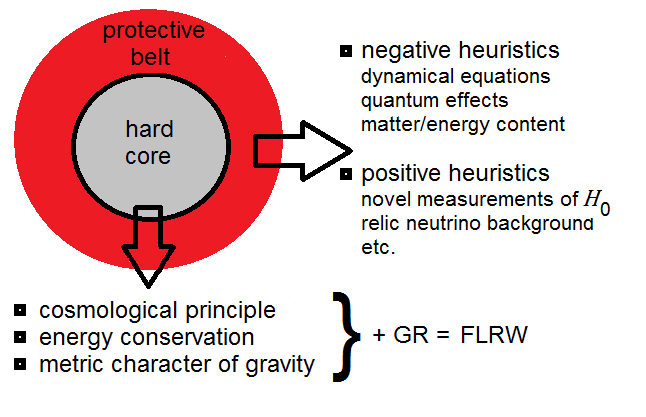}
\caption{A general scheme of Cosmology within the Lakatos' proposal. All the suggested elements are discussed in the text (Section 2)}
\label{fig:cint}
\end{figure}

\subsection{Heuristics in the protective belt}

Lakatos' view of the structure of research programmes envisaged a protective belt around the hard core, with a negative and positive heuristics. Even if his initial definition of "heuristics" changed over the years (Kiss 2006), the prevailing late concept can be stated as saying that "heuristics" studies the patterns of thinking, (to be later related to methodology). Lakatos suggests the existence of a negative heuristics, intended to protect the hard core from being attacked/eroded by the presence of anomalies, redirecting the attention to auxiliary hypothesis and initial conditions. Changes can be made in the latter without affecting the inviolable hard core, held as fundamental for the programme existence. On the other hand, positive heuristics is a set of recommendations articulated to specify how modifications should be performed, including new measurements, verification of predictions and related issues. 

Modern $\Lambda$CDM Cosmology, as any other research programme, has some elements which are identified as integrating the negative heuristics. The first recognition is that Dark Matter and Dark Energy are hypothetical components postulated to avoid a breakdown of the three fundamental points of the hard core, as discussed in the last subsection. This qualifies all forms DM/DE as integrating the negative heuristics, independently of their specific form. In other words, if we could prove that they do not exist, some other element would be necessary to square FLRW models with observations, otherwise it would be necessary to accept that the $\Lambda$CDM framework does not provide a good description of the whole set of data and must be abandoned. 
Much in the same way, modifications to the gravitational theory (for example, introducing a fundamental length) are mainly elements of the negative heuristics, because they would allow to retain the hard core if the modifications fit into a metric framework. 
Other effects entangled with $\Lambda$CDM Cosmology may qualify as negative heuristics elements, and even boost the overall description by "solving" the DM/DE quandary, for instance the presence of quantum effects in the equations, generally neglected but are undoubtedly part of the contemporary physical picture. A full list of candidate elements in the negative heuristics class would be impossible, but the above statements suffice to show some important ones and how they qualify as such.

On the other hand, positive heuristics are devised to improve and move the programme forward, suggesting new developments and measurements. One of the most important of this class is the measurement of the expansion rate $H_0$, a source of concern lately (see below). The "cleanest" suggestion to do so is by means of the standard siren technique (You et al. 2020), which relies on the amplitude of gravitational waves detected in events of mergers of neutron stars and black holes. The novel technique is considered quite free of usual astronomical problems, since gravitational waves are not subject to them, but the number of events and the refinement of the fusion events are still short of providing an accurate ($\sim \, 1 \%$ or so) determination of $H_0$ as needed. 
In summary, there are elements belonging to the protective belt in modern Cosmology which are clearly identified, but an exhaustive list is not possible because of the variety of matters and the enormous amount of work by cosmologists. 

\section{Actual problems (anomalies) arising}

To begin with the anomaly problem, we have addressed DM and DE in some length in a previous article (Horvath 2009), and suggested that these are, by far, the most important anomalies to be solved within $\Lambda$CDM Cosmology. This is also a prime philosophical issue, since even if detected and confirmed, this would mean that $\sim \, ~95 \%$ of the content of the Universe is not what we thought it was. Moreover, in a large class of extra-dimensional theories (Brane World), there is a solution in which  DM and DE are nothing but projections of the (bulk) extra content onto our (brane) Universe. From the point of view of the programme, this can be seen as the injection of external components, which were unpredicted and unexpected, changing the situation quite radically. On the other hand, if DM/DE arise just from a "dark sector" of the Standard Model of fundamental interactions, or even are dark astronomical objects + zero point energy (the latter somewhat strongly suppressed respect to the naive calculation, Zel'dovich 1968), the modern Cosmology programme should go ahead without major problems. This is why initiatives to determine whether DE varies with time through galaxy counts (https://www.darkenergysurvey.org/) are under vigorous development.

An increasingly debated anomaly in Cosmology is the so-called Hubble tension. This issue is simply understood by noting that local determinations of $H_0$ differ from large-scale ones (CMBR and related) by $\geq 4 \sigma$, i.e. much more than the uncertainty bars. In other words, even though expected to yield the same rate $H_0$, different methods point out to incompatible results. A lot of activity has been seen on this serious discrepancy, and its solution may include new physics, with strong consequences for the whole Cosmology. However, in the absence of a hint of solution, it is difficult to forecast what is actually implied for the research programme.

\begin{figure}[htbp]
\centering
\includegraphics[width=8cm]{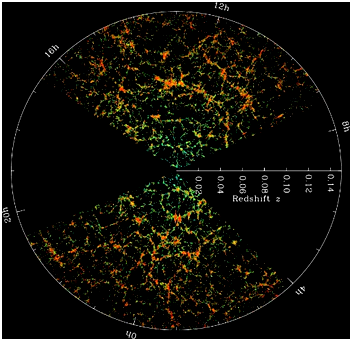}
\caption{An image from the SDSS survey containing 13 billion galaxies (Abazajian et al. 2004). The large-scale distribution of galaxies. This image contains galaxies of all types, which assemble into clusters, superclusters and filaments easily visible to the naked eye. The survey sees an approximate fraction of 1/20th of the currently observable Universe. }
\label{fig:Surv}
\end{figure}

On the other hand, it is known that far beyond the scale of galaxies, a variety of filaments, voids and other inhomogeneities were detected in the largest surveys performed so far (Fig. 2). Some of them are visible to the naked eye, and their evolution can be studied using numerical simulations. According to the idea of homogeneity and isotropy, the structure present at the largest scales must completely “dilute” long before the Hubble scale $\sim 5000 \, Mpc$. However, there are works that have discussed the extent to which this is true in the data, and proposed that the structure does not end, being self-similar up to the limits of the sample. Many of these are fractal models, where the structure features scale invariance because it is a product of some self-organization mechanism. Orthodox cosmologists dismiss a scale-invariant structure all the way up to the Hubble radius, and insist that homogeneity is achieved beyond $\sim 300 \, Mpc$ or so, although much larger structures seem to be present in the data. 

A prime example of these totally unexpected structures that has been identified is the Hercules-Corona Borealis "Great Wall of Quasars" (not to be confused with the original Great Wall which is a "little wall" 10 times smaller in comparison), at scales of $\sim 3000 \, Mpc$ (Fig. 3). Such an enormous structure was not considered possible, and in fact the Cosmological Principle itself and the treatment of the current Universe as a homogeneous and isotropic fluid over a length scale may be now questioned, since this type of structure occupies about half the radius of Hubble. In order to have a consistent perspective with the discussion, we must also remember that to present a self-similar structure or periodicity a physical mechanism to produce perturbations that lead to this structure should exist and be identified, which is not the case yet.

\begin{figure}[htbp]
\centering
\includegraphics[width=8cm]{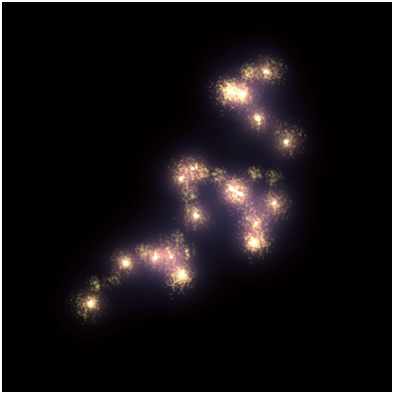}
\caption{An image of the Great Wall of Hercules-Corona Borealis, an association of quasars at a scale comparable to that of the observed Universe (Horv\'ath et al. 2015). The assumption of isotropy and homogeneity of the Cosmological Principle is contested based on this type of structure ``which could not exist''.}
\label{fig:GWQ}
\end{figure}

Another very puzzling feature related to the accepted structure formation has been recently very provided by observations of very early galaxies with large masses (Hainline 2023). The standard scenario of fluctuation growth would not leave enough time to form massive galaxies (with $M \geq 10^{10} \, M_{\odot}$) present up to a redshift $z \sim 15$, that is, only $sim 300-500 \, Myr$ after the Big Bang. On theoretical grounds, much smaller "galactic blocks" with $M \sim 10^{6} \, M_{\odot}$ were expected. Therefore, it is not clear to what extent the view of initially small galaxies, growing a factor $10{3}-10^{4}$ by merging is still tenable or must be revised thoroughly.
The list of anomalies do not stop here: there is a claimed evidence that the dipole asymmetries measured in the CMBR and the ones inferred from matter distributions (Active Galactic Nuclei for instance) do not agree (Singal 2023). It is not possible to make them agree just by adjusting the peculiar motion of the Solar System, therefore, the possibility that both cosmic reference frames are different remains. It is not clear what relationship could have with the other anomalies questioning the Cosmological Principle.

Can these observations, isolated or joint, falsify the Cosmological Principle? the tentative answer is ``yes'', but cosmologists may rescue it by redefining its validity statistically, for example. Anisotropic models of the Universe evolution have existed for decades, and therefore this is not a matter of description, but rather of principles. Thus, the study of structure in the Universe needs to continue and resolve these and other questions of great cosmological interest. 

\section{The Multiverse and its philosophical and scientific unresolved issues }

We have insisted on the role of the Cosmological Principle as an element of the hard core of $\Lambda$CDM, but an opposing view (called the Anthropic Principle) has been discussed to justify our very existence in the late Universe. However, it has been pointed out, and discussed in length, that the Anthropic Principle collides with the Cosmological Principle to some extent. In fact, Carter's (1974) original motivation in the initial formulation of the Anthropic Principle was related, in his opinion, to a degree of overstatement of the power of the latter. Life on Earth and other planets do need special physical conditions, and a place and time for life emergence which is not arbitrary. For example, the fluctuations that led to the formation of the structure can be considered from the point of view of the Anthropic Principle. This has been emphasized by M. Rees (2001), who observed that if the amplitude had been a slightly smaller, the primordial gas would never condense into connected structures, and thus the material enriched in heavy elements by the stars would be dispersed in space and would not allow a later chemical evolution, with a sequence of stellar generations. Now, if the amplitude had been a little higher, regions much larger than clusters of galaxies would have formed very early in the history of the Universe, and would not have broken up into stars, but would have formed vast black holes instead. The remaining gas would be heated to such temperatures that it would emit X-rays and gamma rays, such that material enriched in heavy elements by stars would be quickly trapped in black holes (none of this would favor a Universe where life could flourish to discuss these possibilities today). Therefore, a reasonable "quite" place within a galaxy, and enough time for the planet to evolve from pre-biotic Chemistry to the present state are needed.

A further development has attracted the attention of cosmologists lately, namely the ``Anthropic'' idea that ours is just a favorable Universe out of a set of universes integrating a Multiverse. These alternative Universes, with values of the physical constants and histories that differ from ours, are considered to be real, not a Gibbs ensemble. It has been recognized that this kind of idea brings an epistemic shift to what we consider as "Science", and lead to dead ends in the cognition of the Cosmos (Kragh 2017).

The question is whether an epistemic shift would be acceptable, if ultimately executed. Such attitudes have been seen over the centuries, for example, Cartesians blamed Newtonians for the introduction of unspecified, mystical elements, such as a ``force'' into Natural Philosophy, but our own perception three centuries later does not see anything wrong with it. We may state that an epistemic shift happened 300 years ago, and was completed by successive generations of physicists. Should we "wait and see", or rather comply against the consideration of a physical world which may be unobservable by its very nature? We may be in the middle of such a process. 

\begin{figure}[htbp]
\centering
\includegraphics[width=8cm]{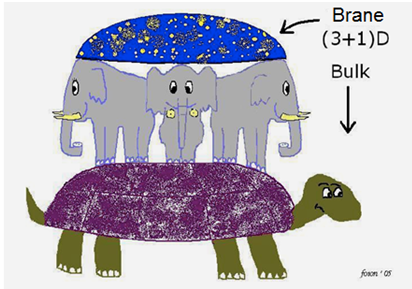}
\caption{Back to square one in Cosmology? A cartoon suggesting the danger of substantial epistemic shifts}
\label{fig:World}
\end{figure}

 \section{Conclusions}
 We have constructed in this paper a picture of the contemporary Cosmology research program patterned after the work and concepts introduced by Imre Lakatos (1978). The research program features a hard core, a negative heuristics and a positive heuristics in a protective belt, identified according to Lakatos' criteria as showed in Section 2. Several anomalies, old and new ones, have been discovered and are under scrutiny at present, and as in other case studies, it is impossible to forecast what the outcome will be when solved.
 
We should remark that to go ahead with this research programme, for many practitioners it is not just more observations what are needed, but rather some general principle(s) based of symmetries, which play the role of justification of cosmological facts. For example, the conservation of angular momentum (as a consequence of the rotation invariance of the physical problem) added to the hypothesis that gravitational forces stem from a central field, were enough to show that planetary orbits should be elliptical, not circular, in the 17th century. There are no such reasons at hand for contemporary Cosmology. For example, a very small, but non-zero cosmological constant, far off the predicted value of Quantum Field Theory (Zel'dovich 1968) does not have a physical reason behind. Other cosmological issues can be thought along the same lines, but for now some problems are introduced just as epicycles of the new picture of the Universe, not necessary features dictated by powerful reasons.

\section{References}

Abazajian, K.  et al. (2004) The Second Data Release of the Sloan Digital Sky Survey. Astronomical Journal 128, 502 

Baez, J. (2020) https://math.ucr.edu/home/baez/noether.html

Beisbart, C. and Jung, T. (2006) Privileged, Typical, or not even that? – our place in the world according to the Copernican and the Cosmological Principles. Journal for General Philosophy of Science 37,225–256

Bellucci, S.,Faraoni, V, Luongo, O. (2022)   Report on session AT7 of the 15th Marcel Grossmann Meeting — "Theories of gravity: alternatives to the cosmological and particle standard models" . The Fifteenth Marcel Grossmann Meeting on General Relativity. Eds. Battistelli, B.S., Jantzen, R.T. and Ruffini, R.. Published by World Scientific Publishing Co. Pte. Ltd., 803-810

Bothe, W. and Geiger, H. (1925) ExperimentelleszurTheorie von Bohr, Kramers und Slater, Die Naturwissenschaften 13, 440–441 

Burnet, J. Early Greek Philosophy. CreateSpace Independent Publishing Platform, USA, 2014.

Carter, B. (1974) Large Number Coincidences and the Anthropic Principle in Cosmology, in: Confrontation of Cosmological Theories with Observational Data, Ed. Longair, M.S. Reidel, Dordrecht, 291

Hainline, K. et al. (2023) The Cosmos in its Infancy: JADES Galaxy Candidates at z > 8 in GOODS-S and GOODS-N.  arXiv:2306.02468

Horvath, J.E. (2009) Dark Matter, Dark Energy and Modern Cosmology: The Case For a Kuhnian Paradigm Shift. Cosmos \& History 5(2), 287–303

Horv\'ath, I., Bagoly, Z., Hakkila, J., Tóth, L. V. (2015). New data support the existence of the Hercules-Corona Borealis Great Wall. Astronomy \& Astrophysics. 584,  A48.

Hu, Jian-Ping and Wang,  Fa-Yin (2023). Hubble Tension: The Evidence of New Physics. Universe 9, 94-136

Kiss, O. (2006) Heuristic, Methodology or Logic of Discovery? Lakatos on Patterns of Thinking. Perspectives on Science 14, 302

Kragh, H..Cosmology and Controversy: The Historical Development of Two Theories of the Universe. Princeton University Press, USA, 1999

Kragh, H. (2017) Fundamental Theories and Epistemic Shifts: Can History of Science Serve as a Guide? astro-ph/ 1702.05648. 

Kramer, M. et al. (2021) Strong-Field Gravity Tests with the Double Pulsar. Physical Review X 11, 041050

Kuhn, T. (1962) The Structure of Scientific Revolutions, University of Chicago Press, USA.

Lakatos, I. The methodology of scientific research programmes. Eds. J. Worrall and G. Currie. Cambridge University Press, UK, 1978.

Landau, L.D.  (1932) On the theory of stars. Phys. Z. Sowjetunion,1, 285-290;   reprinted in Neutron Stars, Black Holes and Binary X-ray Sources, Eds. Gursky, H., Ruffini, R. Reidel, Dordrecht, 1974

Lucretius, De Rerum Natura, 1. 148–156. Legare Street Press, UK, 2023

Oks, E. (2021) Brief review of recent advances in understanding dark matter and dark energy. New Astronomy Reviews 93, 101632

Rees, M. (2001) Just Six Numbers: The Deep Forces That Shape the Universe. Basic Books, USA 
Shutt, T. (2013) Review of Direct Detection of Dark Matter.  American Physical Society, APS April Meeting, April 13-16.

Singal, A.K. (2023) Discordance of dipole asymmetries seen in recent large radio surveys with the Cosmological Principle. arXiv:2303.05141 

You, Z.-Q. et al. (2020) Standard-siren cosmology using gravitational waves from binary black holes. Astrophys. J. 908, id.215

Zel'dovich, Ya. B. (1968) The cosmological constant and the theory of elementary particles. Soviet Physics Uspekhi11, 381

Zwicky, F. (1933). Die Rotverschiebung von extragalaktischen Nebeln. Helvetica Physica Acta 6, 110-127

\end{document}